\documentclass[aps,prB,preprint,superscriptaddress]{revtex4-1}
\usepackage{amsmath,amssymb}
\usepackage{graphicx}
\usepackage{color}
\usepackage{ulem}
\draft

\begin{document}
\makeatletter 
\makeatother

\title{Scalable Parallel Single-Electron Pumps in Silicon with Split-Source Control in the Nanoampere Regime}
\author{Gento Yamahata}
\email{E-mail: gento.yamahata@ntt.com}
\affiliation{NTT Basic Research Laboratories, NTT Corporation, 3-1 Morinosato Wakamiya, Atsugi, Kanagawa 243-0198, Japan}

\author{Takase Shimizu}
\affiliation{NTT Basic Research Laboratories, NTT Corporation, 3-1 Morinosato Wakamiya, Atsugi, Kanagawa 243-0198, Japan}

\author{Katsuhiko Nishiguchi}
\affiliation{NTT Basic Research Laboratories, NTT Corporation, 3-1 Morinosato Wakamiya, Atsugi, Kanagawa 243-0198, Japan}

\author{Akira Fujiwara}
\affiliation{NTT Basic Research Laboratories, NTT Corporation, 3-1 Morinosato Wakamiya, Atsugi, Kanagawa 243-0198, Japan}
\begin{abstract}
Parallelizing single-electron pumps offers a promising route to achieving nanoampere-level currents crucial for quantum current standard applications. Achieving such current levels is essential for demonstrating the ultra-high accuracy of single-electron pumps below 0.1 ppm toward quantum metrology triangle experiments. In addition, improving the accuracy at this current range is also desirable for practical small-current measurements. However, nanoampere-level currents have not yet been achieved with parallel pumps, mainly due to challenges in optimizing operating conditions. Here, we propose a scalable and easily implementable parallelization method based on tunable-barrier single-electron pumps with split source electrodes. By tuning the source voltages, we successfully parallelize four single-electron pumps at 200 MHz and further demonstrate a current plateau exceeding 2 nA using three pumps at 2.1 GHz. The wide applicability of this parallelization technique opens a path toward advancing high-accuracy quantum current standards.
\end{abstract}

\maketitle

\clearpage

%\section*{Introduction}
A tunable-barrier single-electron pump (TSEP) can precisely transfer one electron per cycle under clock control \cite{tunable-barrier1}, making it promising as an accurate current source for quantum current standards \cite{pekola-rev} as well as an on-demand single-electron emitter for quantum information processing \cite{SEsouce1,GYnnano,Jcol,Ncol}. Scaling up the number of TSEPs is important for both applications, and for quantum current standards, parallelization is a practical approach to increase current output.

The importance of increasing the output current lies in its connection to the accuracy of current generation, which, in turn, directly affects the accuracy of current measurements. In general, as the current level decreases below 1 \(\mu\)A, the accuracy of current measurements deteriorates \cite{LNECCC,PQCS}. Therefore, it becomes essential to improve the precision of small current generation across a wide current range. This need is also underscored by the growing demand for ultralow current measurements with high accuracy for radiation measurement, air pollutant detection, and medical applications \cite{AISTrev}.

Efforts to address this need have focused on developing new types of current sources, aiming to extend high-precision current generation beyond the capabilities of conventional standards. A quantum current source based on the quantum Hall and the Josephson effects enables ultra-high-precision current generation with uncertainties around 0.01 ppm in the milliampere and microampere ranges \cite{LNECCC,PQCS}. However, extending this level of precision to the nanoampere regime remains a significant challenge. In contrast, TSEPs operating at a clock frequency \( f \) can generate an accurate small current given by \(e f \), where \( e \) is the elementary charge \cite{tunable-barrier1}. Previous studies on TSEPs based on GaAs or silicon have demonstrated error rates of 0.1 to 1 ppm or lower in the picoampere regime, with the error rates being limited by measurement uncertainties \cite{gib1,PTB-ulca1,Kriss-HR,NPL-NTT1, Zhao_pump,RobPTB,KrissAIST,LHeSi,Si2GHz}. As a natural extension, increasing the pumped current to reach the nanoampere regime is a crucial next step. In addition, higher-accuracy operation of TSEPs in the nanoampere regime is also important for more precise comparisons of fundamental physical constants in the quantum metrology triangle experiment \cite{QMT_first,capNIST}.

There are two primary approaches to increasing the output current of TSEPs. The first approach is to increase the operating frequency \(f\). Indeed, high-precision measurements have been reported in the gigahertz regime \cite{UnivRev}. A key advantage of TSEPs is their ability to maintain high accuracy in this frequency range. However, the present upper limit is around 2 GHz for TSEPs that transfer electrons through an electrically defined quantum dot (QD) \cite{Si2GHz}. Beyond this frequency, non-adiabatic excitations become significant \cite{kataoka1}, leading to a substantial deterioration in accuracy. On the other hand, in TSEPs that transfer electrons via trap levels \cite{trap1}, operation at 7.4 GHz has been reported to generate nanoampere-level currents \cite{trap2}. Yet, precise position control of the trap levels is difficult, making it challenging to fabricate multiple identical devices, which poses a reproducibility issue.

Due to the challenges in increasing the operating frequency, the second approach of parallelizing multiple TSEPs becomes particularly important. In the simplest configuration, all TSEPs share the two gate electrodes used to form the entrance and exit barriers of the QD, as well as the source and drain electrodes, making it suitable for large-scale integration. However, since the optimal conditions for accurate current generation differ among individual TSEPs \cite{10.1063/5.0225998}, achieving high-precision operation in such a fully shared configuration is not feasible. To enable individual voltage tuning, several approaches have been proposed, such as wiring each TSEP separately \cite{PhysRevB.80.113303, doi:10.1021/acs.nanolett.3c02858}, adding extra gate electrodes \cite{BaePara}---as implemented in parallel SINIS turnstiles \cite{PekolaPara,naka1}---and implementing individually addressable exit gate electrodes \cite{PTBpara}. Nevertheless, no previous studies have demonstrated the generation of currents exceeding the nanoampere level with precise and independent control in a parallelized TSEP system.

Here, we propose parallel TSEPs with individual source electrodes as a simple and practical configuration for independent TSEP tuning. Controlling individual source voltages enables selective activation of TSEPs and precise optimization of their operating conditions. Using this method, we successfully tuned multiple TSEPs to their optimal operating points and demonstrated nanoampere-level operation exceeding 2 nA.

%\section*{Results}
%\subsection*{Proposed Reconfigurable Parallelization Method}
We fabricated a device consisting of eight silicon TSEPs arranged in parallel (Fig.~1a). The fabrication process is described in Supporting Information. Each TSEP shares entrance gate G\(_{\mathrm{ent}}\) and exit gate G\(_{\mathrm{exit}}\), which have widened sections between adjacent TSEPs to support structural stability. A QD is electrically formed in the silicon nanowire between the entrance and exit gates of each TSEP (Fig.~1b). The device has a single drain electrode, while eight source electrodes (S\(n \): \(n=1,2,\dots,8\)) are individually connected to each TSEP.

Electrons are induced by applying DC voltages to an additional upper gate, which covers the entire TSEP region, and a substrate gate. A high-frequency voltage \(V_{\mathrm{ent}}^{\mathrm{ac}}(t)\) and a DC voltage \(V_{\mathrm{ent}}\) are applied to G\(_{\mathrm{ent}}\), while a DC voltage \( V_{\mathrm{exit}}\) is applied to G\(_{\mathrm{exit}}\). Source voltages \(V_{\mathrm{S}n} \) are independently controlled by separate voltage sources---an essential feature of our proposal. The pumped current \( I_{\mathrm{pump}}\) is generated and measured using an ammeter at the drain side.

The operating principle of the TSEP is illustrated in Fig.~1c \cite{kaestner1, GYgval}, which shows potential diagrams ordered in time to depict the changes during single-electron transfer. When the entrance barrier is sufficiently low, electrons are loaded from the source S\( n\). Subsequently, due to the capacitive coupling between G\(_{\mathrm{ent}}\) and the QD, the potential of the QD rises as the entrance barrier rises. After the highest occupied energy level of the QD with \(m\) electrons aligns with the source Fermi level \(E_\mathrm{F}(V_{\mathrm{S}n})\), an electron can tunnel back to the source at an escape rate \( \Gamma_{m} \). Here, \(V_{\mathrm{S}n}\) is explicitly written as the argument of \(E_\mathrm{F}\), emphasizing that the Fermi level is controlled via the source voltage. However, in the case where the entrance barrier becomes sufficiently high, the escape rate \( \Gamma_{m} \) becomes small, leading to the capture of \(m\) electrons in the QD (Fig.~1c illustrates the case of single-electron capture). We refer to this process as dynamic capture. Finally, when the QD potential becomes sufficiently high, the captured electrons are ejected into the drain. When \(m\) electrons are transferred per cycle, the pumped current is given by \(I_{\mathrm{pump}} = m e f\). The current accuracy is mainly determined by the accuracy in the dynamic capture process because the loading and ejection errors can be minimized via gate voltage control.

In this work, we optimize the dynamic electron capture by adjusting the source voltage. Using the potential diagram in Fig.~1d, we explain how an increase in the source voltage decreases electron capture probability. Consider the situation where the single-electron energy level of the QD aligns with the Fermi level \(E_\mathrm{F}(0)\) at time \( \tau_{0}\), with electrons tunneling back to the source at an escape rate \( \Gamma_{1} \). When a voltage \(V \) is applied to the source, the Fermi level shifts downward by an amount of \(eV \). In that case, we need to consider an earlier time \( \tau_{V}\) (\( \tau_{V} < \tau_{0}\)), at which the time-dependent QD energy level becomes aligned with the Fermi level \(E_\mathrm{F}(V)\), without changing any other DC voltages. Since the entrance gate is capacitively coupled most strongly to the entrance barrier, the energy difference between the QD potential and the barrier top at \( \tau_{V}\) is smaller than that at \( \tau_{0}\), leading to a larger escape rate at \( \tau_{V}\). This, in turn, decreases the dynamic electron capture probability. Note that when \(V\) is sufficiently large, electrons always tunnel back to the source and the loading process is also hindered, disabling the TSEP.

%\subsection*{Experimental Demonstration of Reconfigurable Parallelization}
We then experimentally demonstrate the parallelization method (Fig.~2). Figure~2a shows a two-dimensional plot of \(I_{\mathrm{pump}}\) as a function of \(V_{\mathrm{S8}}\) and \(V_{\mathrm{exit}}\), with several labels indicating current plateau values shown in red. The current is generated only from the TSEP connected to S8, with the other TSEPs disabled by applying positive biases to the sources. The effect of tuning the dynamic capture condition via \(V_{\mathrm{S8}}\) is visible as the boundary indicated by the slanted black dashed line in Fig.~2a. An analytical expression for this boundary is also provided in Supporting Information. On the other hand, a horizontal boundary observed in Fig.~2a is independent of \(V_{\mathrm{S8}}\). The boundary is determined by the ejection process because the QD is effectively decoupled from the source in this process, owing to the entrance barrier being sufficiently high and wide.

Subsequently, the TSEP connected to S2 is activated in addition to the one connected to S8, resulting in the two-dimensional current map shown in Fig.~2b. In the region where \(V_{\mathrm{S8}} > 0\), only the TSEP connected to S2 is active, and labels indicating its current plateau values are shown in black. Since this TSEP is independent of variations in \(V_{\mathrm{S8}}\), the boundaries of the current plateaus remain parallel to the horizontal axis. In the region where \(V_{\mathrm{S8}} < 0\), the combined current from both TSEPs appears.

Figure 2c shows the \(V_{\mathrm{S8}}\) dependence of \( I_{\mathrm{pump}}\), where blue and red curves correspond to the blue and red dashed lines in Fig.~2b. The blue data represent the characteristics of only the TSEP connected to S8. In contrast, the red data reflect the combined operation of the two TSEPs, with the contribution from the TSEP connected to S2 shifting the total current upward by \(ef\). Since a TSEP typically exhibits high accuracy in the \(ef\) plateau, the precision of the \(ef + ef\) plateau achieved by two TSEPs combined is expected to be better than that of a \(2ef\) plateau from a single TSEP.

We now describe in detail the method for tuning the parallel TSEPs to their optimal operating conditions (Figs.~2d and 2e). A practical approach to finding the optimal condition for the first TSEP (the one connected to S2 in the present case) is to plot the logarithm of the absolute value of the deviation of the normalized current from the ideal value of 1 as a function of \(V_{\mathrm{exit}}\) (Fig.~2d) \cite{LHeSi,GYgval,holeJAP,10.1063/5.0231792}. The optimal point is the intersection of the extended linear fits to the left- and right-sloping parts of the data. After fixing \(V_{\mathrm{exit}}\), the second TSEP (the one connected to S8 in the present case) is activated by sweeping the source voltage, as shown in Fig.~2e. The same method is applied to the deviation of the normalized current from the ideal value of 2, to determine the optimal source voltage. By repeating this process sequentially for the source voltages of other TSEPs, the optimal source voltages for each TSEP are identified. It is important to note that the overall accuracy of the parallel TSEPs is limited by the least accurate TSEP. In this experiment, the accuracy of the two parallel TSEPs is approximately \(10^{-3}\), constrained by the first TSEP, although the second TSEP achieves an accuracy of about \(10^{-5}\). 

%\subsection*{Optimized Parallelization of Four TSEPs}
As a proof of concept for scalable TSEP parallelization, we demonstrate the operation of up to four TSEPs in parallel (Fig.~3). Figure 3a shows a two-dimensional plot of \(I_{\mathrm{pump}}\) as a function of \(V_{\mathrm{S7}}\) and \(V_{\mathrm{exit}}\). In this plot, the third TSEP connected to S7 is activated in addition to the previous two TSEPs, while keeping \(V_{\mathrm{S8}}\) fixed at the black dashed line in Fig.~2b. The blue labels indicate the value of the current generated by the third TSEP. The ejection boundary of the TSEP connected to S7 (green dashed line) coincidentally aligns with the onset of the current from the TSEP connected to S8 (blue dashed line). Figure 3b displays \(I_{\mathrm{pump}}\) along the red dashed line in Fig.~3a, demonstrating the generation of a \(3ef\) current plateau as the sum of the \(ef\) current plateaus from each of the three TSEPs.

With \(V_\mathrm{S7}\) fixed at the black dashed line in Fig.~3a, the current generated from the fourth TSEP connected to S3 is added (Fig.~3c). The green label indicates the value of the current generated by the fourth TSEP. In this case, the ejection boundary (green dashed line) appears inside the \(2ef\) plateau of the previous two TSEPs connected to S7 and S8. As seen in Figs.~3a and 3c, the ejection condition slightly varies for each TSEP. However, since the exit barrier can be lowered by setting \(V_{\mathrm{exit}}\) to a less negative value, the optimal condition for parallel operation can be obtained away from the ejection boundary. Figure 3d displays \(I_{\mathrm{pump}}\) along the red dashed line in Fig.~3c, demonstrating the generation of a \(4ef\) current plateau as the sum of the \(ef\) current plateaus generated from each of the four TSEPs. Since this process can be repeated as many times as needed, the limitation of the parallelization is only the number of TSEPs that can be integrated at once.

%\subsection*{Achieving Nanoampere Currents by Parallelizing TSEPs}
Using the parallelization method demonstrated above, we explored high-current operation (Fig.~4). To increase the current generated from an individual TSEP, we set the frequency to 2.1 GHz. While higher frequencies generally degrade pumping accuracy, three of the four TSEPs maintained stable current plateaus even at this frequency. Figure~4a shows the parallel operation with the two TSEPs connected to S7 and S8 at 2.1 GHz, demonstrating a current map similar to that observed at low frequency. Figure 4b presents the current-voltage characteristics measured along the red dashed line in Fig.~4a. Note that we can flexibly select not only the \(ef\) current plateau but also higher current plateaus, including \(2ef\), \(3ef\), and beyond, by tuning the source voltage. In this demonstration, we selected the \(ef + 2ef\) plateau, achieving a current exceeding 1 nA.

With \(V_\mathrm{S7}\) fixed at the black dashed line in Fig.~4a, we demonstrate the parallel operation of the three TSEPs at 2.1 GHz (Fig.~4c). Figure 4d presents the current-voltage characteristics measured along the red dashed line in Fig.~4c. Markedly, current plateaus up to \(ef + 2ef + 3ef\) were achieved, leading to current levels exceeding 2 nA—a notable enhancement in current generation using single-electron pumps. This current level corresponds to an operating frequency of 12.6 GHz, a regime that remains highly challenging for TSEPs.

%\section*{Discussion and Conclusion}
In this study, we proposed and demonstrated a method for optimizing operating conditions and achieving large-scale parallelization of TSEPs. Using four parallel TSEPs, we achieved well-defined current generation by stacking four \(ef\) current plateaus. Additionally, the achievement of 2 nA current generation using three parallel TSEPs demonstrates the potential of this approach for enabling high-current operation. An important aspect of the proposed method is its broad applicability to TSEPs employing a split-source electrode configuration, regardless of gate structure, channel configuration, or material. Once combined with sufficiently accurate TSEPs, this approach could lead to high-precision current generation at the nanoampere level. In contrast, other approaches using a superconducting Josephson junction device with dual Shapiro steps \cite{Crescini2023,Kaap2024} or a quantum anomalous Hall device \cite{QAHcurrent} have also been explored for nanoampere current generation. However, none of these devices have demonstrated accuracy below 0.1 ppm. Given this context, the parallelization technique presented here could contribute to realizing a high-precision current standard in the nanoampere regime in the future.

\section*{Supporting Information}
\subsection*{Device Fabrication}
Eight silicon nanowires were fabricated on a silicon-on-insulator substrate with a 400-nm-thick buried oxide layer using electron beam lithography, followed by dry etching. The designed width of the nanowires in this lithography process was 40 nm. After thermal oxidation of the nanowires to form a 30-nm gate insulator, two heavily n-doped polycrystalline silicon lower gate electrodes were fabricated above each nanowire using electron beam lithography and dry etching. The design length of the lower gates was 90 nm, with a designed gate spacing between the two lower gates of 80 nm. The lower gates were thermally oxidized, and an additional 50 nm silicon oxide layer was deposited using chemical vapor deposition. Subsequently, upper gate electrodes, also made of heavily n-doped polycrystalline silicon, were formed using photolithography and dry etching to cover the entire TSEP structure. The upper gates served as a mask for ion implantation, where phosphorus was implanted at high concentrations into the source and drain regions. Finally, aluminum electrodes were deposited by vacuum evaporation, and ohmic contacts to the source and drain were formed using photolithography and wet etching.

\subsection*{Measurement Setup}
The measurements were performed using a dilution refrigerator with a base temperature of 20 mK. All DC voltages applied to the source and gate electrodes were generated using Yokogawa GS210 DC voltage sources. High-frequency signals were supplied as sine waves by a Keysight E8257D analog signal generator. The generated current was converted to a voltage using an NF CA5351 programmable current amplifier, and the voltage was measured with a Keysight 3458A digital multimeter. A 3-dB attenuator was placed on the high-frequency signal line at room temperature. The input power of the high-frequency signal was set to 10 dBm, resulting in an effective power of 7 dBm delivered to the device inside the dilution refrigerator. The line connected to S5 in the refrigerator exhibited a current leak, so S5 was electrically floated at room temperature. This floating configuration was also verified to prevent any unintended current output from the S5 terminal. For TSEPs not used in the experiment (other than the one connected to S5), a source voltage of 0.3 V was applied to disable them. The upper gate voltage was fixed at -0.35 V, and the substrate gate voltage was maintained at 18 V throughout the measurements.

\subsection*{Analytical Expression of the Pump Current}
Following Ref.~33, the analytical expression for the current in the dynamic capture condition as a function of the source voltage in a TSEP is derived. The electron capture probability \(P_1\) for the QD in the TSEP is given by:
$$
P_1 = \int_{\mu_0}^{\infty} D_{\mathrm{ex}}(\mu_1) F(\mu_1) d\mu_1,
$$
where
$$
D_{\mathrm{ex}}(\mu_1) = \exp\left[-\exp\left(-\frac{\mu_1-\mu_1^c}{gkT_0}\right)\right],
\quad F(\mu_1) = -\frac{df(\mu_1)}{d\mu_1},
$$
with the Fermi function \(f(x) = \left[1+\exp(x/kT)\right]^{-1}\). Here, \(k\) is the Boltzmann constant, \(T\) is the temperature, \(T_0\) is the characteristic temperature for tunneling, \(\mu_0\) is the electrochemical potential of the QD ground state at time 0, and \(\mu_1^c\) is the electrochemical potential when detailed balance breaks. The parameter \(g\) represents the ratio of the change in the QD potential to that in the entrance barrier height relative to the QD potential, as \( V_{\mathrm{ent}}\) is varied. When the source voltage \(V_{\mathrm{S}n}\) is applied, the Fermi level shifts by \(eV_{\mathrm{S}n}\). Under strong capacitive coupling between the entrance gate and QD, the Fermi function can be approximated as a delta function \(F(-eV_{\mathrm{S}n}) \approx \delta(-eV_{\mathrm{S}n})\). Consequently, the capture probability simplifies to:
$$
P_1 \approx D_{\mathrm{ex}}(-eV_{\mathrm{S}n}) = \exp\left[-\exp\left(\frac{eV_{\mathrm{S}n}+\mu_1^c}{gkT_0}\right)\right].
$$
By substituting the explicit form of \(\mu_1^c\) from Ref.~33, the capture probability can be further expressed as:

$$
P_1 = \exp\left[-\exp\left(-\frac{\left[(1+\frac{1}{g}) \alpha_\mathrm{e} - \alpha_{\mathrm{eB}}\right] V_\mathrm{GD} + \left[(1+\frac{1}{g}) \alpha_\mathrm{s} - \alpha_{\mathrm{sB}} - \frac{e}{g}\right] V_{\mathrm{S}n} + \left (1+\frac{1}{g}\right ) E_C}{kT_0} +\text{const.}\right)\right].
$$
Here, the source voltage effects on both the QD and the entrance barrier are considered analogous to those of the exit gate voltage. The parameters \(\alpha_\mathrm{e}\) and \(\alpha_{\mathrm{eB}}\) denote the energy change ratios of the QD and the entrance barrier by changing the exit gate voltage, respectively. Similarly, \(\alpha_\mathrm{s}\) and \(\alpha_{\mathrm{sB}}\) correspond to those by changing the source voltage. \(E_C\) is the charging energy of the QD. The slope of the dynamic capture boundary shown in Fig.~2a is determined by the ratio of the prefactors of \(V_{\mathrm{exit}}\) and \(V_{\mathrm{S}n}\) in the exponent of this expression.

%merlin.mbs apsrev4-1.bst 2010-07-25 4.21a (PWD, AO, DPC) hacked
%Control: key (0)
%Control: author (8) initials jnrlst
%Control: editor formatted (1) identically to author
%Control: production of article title (-1) disabled
%Control: page (0) single
%Control: year (1) truncated
%Control: production of eprint (0) enabled
%

%\section*{Contributions}
%G.Y. designed the device and the experiment. G.Y. fabricated the sample with contributions from T.S. and K.N. G.Y. measured and analyzed the data with support from A.F. G.Y. wrote the manuscript with input from all authors.

\clearpage

\begin{figure}
\begin{center}
\includegraphics[pagebox=artbox]{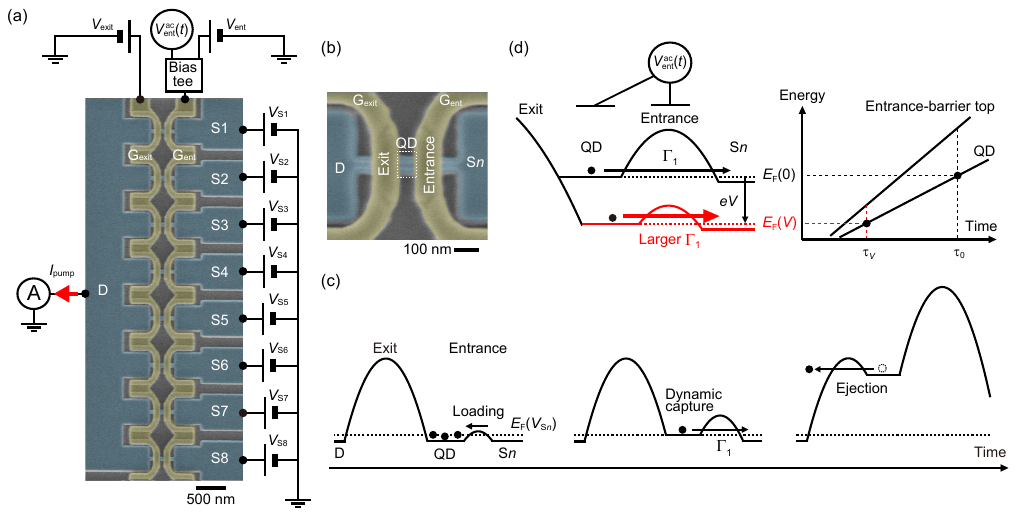}%width=230pt  
\end{center}
\caption{(a) False color scanning electron microscope (SEM) image of the parallel tunable-barrier single-electron pumps (TSEPs) with the electrical measurement setup. (b) False color SEM image of an individual TSEP. (c) Schematic of the electron potentials illustrating the operating principle of the TSEP. The entrance barrier height and quantum-dot (QD) potential increase with increasing time. (d) Schematic of the electron potential illustrating changes in the dynamic capture condition due to source voltage variation, with the corresponding time evolution of energy shown on the right. The capacitive coupling of the high-frequency signal to both the entrance barrier and the QD is also indicated.}
\label{f1}
\end{figure}

\begin{figure}
\begin{center}
\includegraphics[pagebox=artbox]{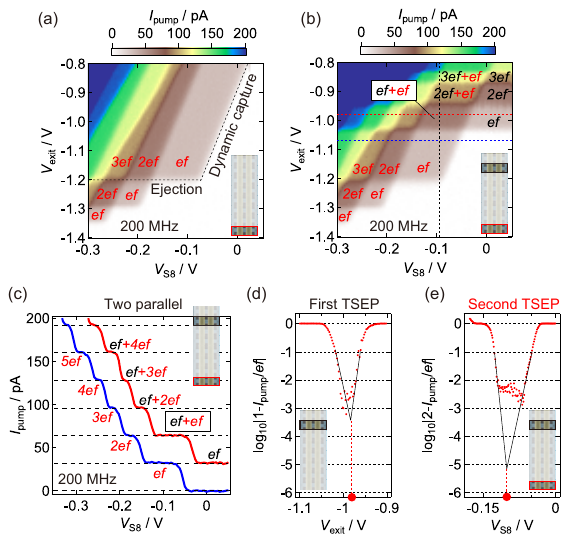}%width=230pt
\end{center}
\caption{(a) Two-dimensional plot of the current \(I_{\mathrm{pump}}\) generated by the TSEP connected to S8 as a function of \(V_{\mathrm{S8}}\) and \(V_{\mathrm{exit}}\), with an operating frequency of 200 MHz and \(V_{\mathrm{ent}} = -1.13\) V. (b) Two-dimensional plot of the current \(I_{\mathrm{pump}}\) generated by the two TSEPs connected to S2 and S8 as a function of \(V_{\mathrm{S8}}\) and \(V_{\mathrm{exit}}\), with an operating frequency of 200 MHz, \(V_{\mathrm{ent}} = -1.13\) V, and \(V_{\mathrm{S2}} = -0.15\) V. (c) Dependence of \(I_{\mathrm{pump}}\) on \(V_{\mathrm{S8}}\) along the red and blue dashed lines in Fig.~2b for \(V_{\mathrm{exit}} = -0.98\,\mathrm{V}\) (red) and \(-1.07\,\mathrm{V}\) (blue), respectively. (d) Plot of \(\log_{10} \left|1 - I_{\mathrm{pump}}/ef \right|\) as a function of \(V_{\mathrm{exit}}\) for the first TSEP connected to S2, with \(V_{\mathrm{S8}} = 0.05\,\mathrm{V}\), corresponding to the right edge of Fig.~2b. (e) Plot of \(\log_{10} \left|2 - I_{\mathrm{pump}}/ef \right|\) as a function of \(V_{\mathrm{S8}}\) for the second TSEP connected to S8, with \(V_{\mathrm{S2}} = -0.15\,\mathrm{V}\) and \(V_{\mathrm{exit}} = -1\,\mathrm{V}\). The insets in (a)–(e) indicate the positions of the TSEPs used in each measurement.}
\label{f2}
\end{figure}

\begin{figure}
\begin{center}
\includegraphics[pagebox=artbox]{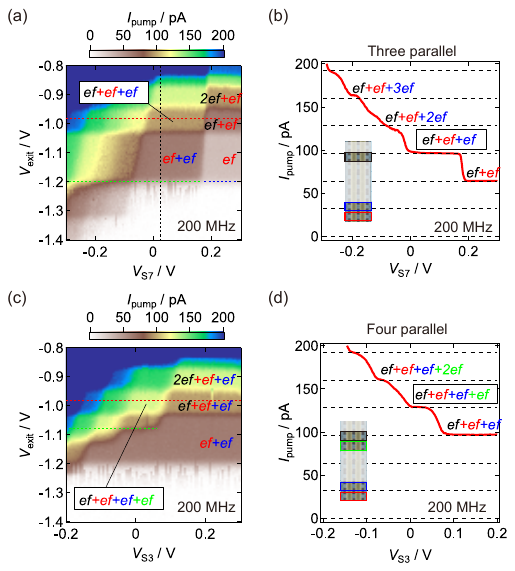}%width=230pt
\end{center}
\caption{(a) Two-dimensional plot of the current \(I_{\mathrm{pump}}\) generated by the three TSEPs connected to S2, S7, and S8 as a function of \(V_{\mathrm{S7}}\) and \(V_{\mathrm{exit}}\), with an operating frequency of 200 MHz, \(V_{\mathrm{ent}} = -1.13\) V, \(V_{\mathrm{S2}} = -0.15\) V, and \(V_{\mathrm{S8}} = -0.095\) V. (b) Dependence of \(I_{\mathrm{pump}}\) on \(V_{\mathrm{S7}}\) along the red dashed line in Fig.~3a, with \(V_{\mathrm{exit}}=-0.98\) V. (c) Two-dimensional plot of the current \(I_{\mathrm{pump}}\) generated by the four TSEPs connected to S2, S3, S7, and S8 as a function of \(V_{\mathrm{S3}}\) and \(V_{\mathrm{exit}}\), with an operating frequency of 200 MHz, \(V_{\mathrm{ent}} = -1.13\) V, \(V_{\mathrm{S2}} = -0.15\) V, \(V_{\mathrm{S8}} = -0.095\) V, and \(V_{\mathrm{S7}} = 0.025\) V. (d) Dependence of \(I_{\mathrm{pump}}\) on \(V_{\mathrm{S3}}\) along the red dashed line in Fig.~3c, with \(V_{\mathrm{exit}}=-0.98\) V. The insets in (b) and (d) indicate the positions of the TSEPs used in each measurement.}
\label{f3}
\end{figure}

\begin{figure}
\begin{center}
\includegraphics[pagebox=artbox]{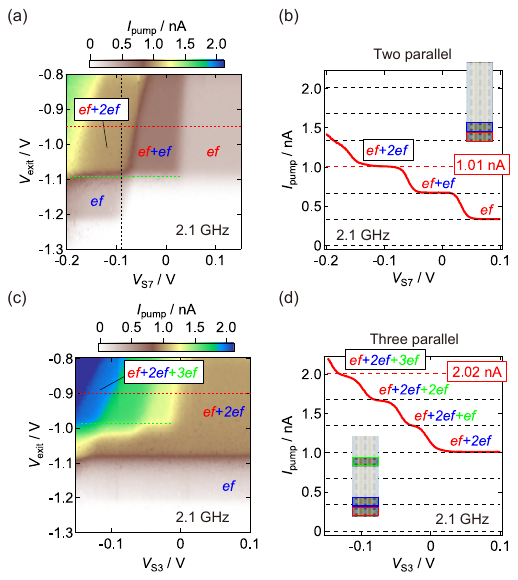}%width=230pt
\end{center}
\caption{(a) Two-dimensional plot of the current \(I_{\mathrm{pump}}\) generated by the two TSEPs connected to S7 and S8 as a function of \(V_{\mathrm{S7}}\) and \(V_{\mathrm{exit}}\), with an operating frequency of 2.1 GHz, \(V_{\mathrm{ent}} = -1.25\) V, and \(V_{\mathrm{S8}} = -0.085\) V. The green dashed line is the ejection boundary. (b) Dependence of \(I_{\mathrm{pump}}\) on \(V_{\mathrm{S7}}\) along the red dashed line in Fig.~4a, with \(V_{\mathrm{exit}} = -0.95\) V. (c) Two-dimensional plot of the current \(I_{\mathrm{pump}}\) generated by the three TSEPs connected to S3, S7, and S8 as a function of \(V_{\mathrm{S3}}\) and \(V_{\mathrm{exit}}\), with an operating frequency of 2.1 GHz, \(V_{\mathrm{ent}} = -1.25\) V, \(V_{\mathrm{S8}} = -0.085\) V, and \(V_{\mathrm{S7}} = -0.09\) V. The green dashed line is the ejection boundary. (d) Dependence of \(I_{\mathrm{pump}}\) on \(V_{\mathrm{S3}}\) along the red dashed line in Fig.~4c, with \(V_{\mathrm{exit}} = -0.9\) V. The insets in (b) and (d) indicate the positions of the TSEPs used in each measurement.}
\label{f4}
\end{figure}

\end{document}